\documentclass[aps,prd,twocolumn,groupedaddress,showpacs]{revtex4}
\usepackage{graphicx}
\usepackage{dcolumn}
\usepackage{bm}
\usepackage{amsmath}
\usepackage{epstopdf}
\usepackage{amsfonts}
\usepackage{amssymb}%

\usepackage[colorlinks=true,citecolor=blue,urlcolor=magenta,breaklinks]{hyperref}

\usepackage{natbib}

\providecommand{\U}[1]{\protect\rule{.1in}{.1in}}

\newcommand{\be}{\begin{equation}}
\newcommand{\en}{\end{equation}}
\newcommand{\bea}{\begin{eqnarray}}
\newcommand{\ena}{\end{eqnarray}}

\begin{document}

\title{The dark matter effect \\ on realistic equation of state in neutron stars}

\author{Grigorios Panotopoulos and Il\'\i dio Lopes}
\email{grigorios.panotopoulos@tecnico.ulisboa.pt, ilidio.lopes@tecnico.ulisboa.pt}
\affiliation{Centro Multidisciplinar de Astrof\'{\i}sica, Instituto Superior T\'ecnico,
Universidade de Lisboa, Av. Rovisco Pais, 1049-001 Lisboa, Portugal}

\date{\today}

\begin{abstract}
In this work we apply relativistic mean-field theory in neutron stars assuming that fermionic dark matter is trapped inside the star and interacts directly with neutrons by exchanging Standard Model Higgs bosons. For realistic values of the parameters of the model we compute numerically the equation of state, and 
we compare it to the standard one. Furthermore, the mass-to-radius relation for both equations of state (pure neutron matter as well as admixed DM-neutron star) is shown, and the highest star mass for both cases is reported.
\end{abstract}


\maketitle

\section{Introduction}

Neutron stars are exciting objects and excellent natural laboratories to test non-standard physics, since from the one hand they are the densest objects in the universe after black holes, and on the other hand studying its properties requires the combination of several areas of physics, from gravity to particle physics to thermodynamics and statistical physics. The gravitational theory provides us with the Tolman-Oppekheimer-Volkoff (TOV) equations \cite{TOV}, while particles physics and statistical physics provide us with the equation of state (EoS) that is also needed to close the system of equations. In modern times realistic equations of state are obtained in the framework of relativistic mean-field theory (RMF) models \cite{walecka1,walecka2} that have become the most used ones. Since they are relativistic, by construction the causality condition is automatically satisfied. Although current data are still limited, the recently observed pulsar PSR J1614-2230 with a mass
$(1.97 \pm 0.04) M_{\odot}$ \cite{pulsar} has put a stringent constrain on the hadronic EoS. On the predicted mass-radius diagram the highest neutron star mass crucially depends on the equation of state, and soft equations of state predict lower highest masses \cite{paper}. Any EoS line that does not intersect the J1614−2230 band is ruled out by this measurement. In particular, according to Fig. 3 of \cite{pulsar}, most EOS curves involving exotic matter, such as kaon condensates or hyperons, tend to predict maximum neutron star masses well below $2 M_{\odot}$, and are therefore ruled out.

Several well-established observational data from Cosmology and Astrophysics strongly suggest that the non-relativistic matter component of the Universe is dominated by a new type of matter particles, yet to be discovered, the so-called Dark Matter (DM). It was in 1933 when Zwicky studying clusters of galaxies introduced the term "missing mass" or dark matter \cite{zwicky}. Much later Rubin and Ford with optical studies of M31 made the case for DM in galaxies in 1970 \cite{rubin}. For a review on dark matter see e.g. \cite{olive,munoz}. Despite the fact
that as of today there are many DM candidates \cite{taoso,Martins2017,LS2014,LS2010,Kouvaris10,britoetal1,britoetal2}, the nature and origin of DM still remains a mystery, comprising one of the biggest challenges in modern theoretical cosmology.

Even if dark matter does not interact directly with normal matter, it can have significant gravitational effects on stellar
objects \cite{massiveNS,chinos,admixed}. In these works it was assumed that normal matter and DM interact only through gravity. However, in many popular models the DM particle, irrespectively of its spin, interacts directly with nucleons by exchanging Standard Model Higgs bosons \cite{portal} (although there is no fundamental reason for that, and in principle the coupling of the DM particle to the Higgs boson could be zero), and the relevant Feynman diagram is precisely the one used in DM direct detection searches \cite{detection1,detection1b,detection2,Jose2016}.

It is the aim of the present work to study the DM effect on the nucleon equation of state in the framework of relativistic mean-field theory taking into account the DM-nucleon interaction and considering realistic values for the Yukawa coupling constants. Our work is organized as follows: In the next Section we briefly summarize
RMF, while in Section 3 we study the DM effect and we compare with the standard equation of state in pure nuclear matter without dark matter. Finally we summarize our work in the final Section. We work in natural unites where $c=\hbar=1$.

\section{Relativistic mean-field theory}

In this Section we briefly summarize the basic ingredients of RMF models \cite{walecka1,walecka2}, and for a recent discussion see e.g. \cite{basic}.
We employ the simplest version of quantum hydrodynamics, also known as $\sigma-\omega$ model (for more elaborated models see \cite{model1,model2,model3,model4}), in which the strong force between neutrons is realized by the exchange of a scalar meson $\phi$ responsible for the strong attractive force, and the exchange of a vector meson $V_\mu$ responsible for the strong repulsive force. The system is described by the hadronic Lagrangian density
\begin{widetext}
\be
\begin{split}
\mathcal{L}_{had}=\bar{\psi} (i \gamma_\mu \partial^\mu-m_N+g_s \phi+g_v \gamma^\mu V_\mu) \psi+\frac{1}{2} (\partial_\mu \phi \partial^\mu \phi-m_s^2 \phi^2)-\frac{1}{4}V_{\mu \nu} V^{\mu \nu}+\frac{1}{2} m_\omega V_\mu V^\mu
\end{split}
\en
\end{widetext}
where $V_{\mu \nu}=\partial_\mu V_\nu-\partial_\nu V_\mu$ is the field strength of the vector meson.
The masses $m_s,m_\omega$ and the couplings $g_s,g_v$ take the numerical values \cite{serot,thesis}
\begin{eqnarray}
m_s & = & 520 \: MeV \\
m_\omega & = & 783 \: MeV \\
g_s^2 & = & 109.6 \\
g_v^2 & = & 190.4
\end{eqnarray}
while the neutron mass is taken to be $m_N \simeq 1 GeV$.
In the mean-field approximation it is assumed that both meson fields are constants $\phi_0,V_0$, and therefore the system looks like an ideal fermion gas where neutrons acquire an effective mass \cite{thesis}
\be
m_*=m_N-g_s \phi_0
\en
Since the kinetic terms for the mesons vanish the total pressure and energy density of the system is given by \cite{thesis}
\begin{eqnarray}
p & = & p_{st}(m_*) - \frac{m_s^2 \phi_0^2}{2} + \frac{m_\omega^2 V_0^2}{2} \\
\epsilon & = & \epsilon_{st}(m_*) + \frac{m_s^2 \phi_0^2}{2} + \frac{m_\omega^2 V_0^2}{2}
\end{eqnarray}
where $p_{st}(m_*),\epsilon_{st}(m_*)$ are the standard expressions for the pressure and the energy density respectively of an ideal Fermi gas evaluated at the effective mass $m_*$. 
They are given by \cite{basic}

\begin{eqnarray}
\epsilon_{st} & = & \frac{2}{(2 \pi)^3} \int_0^{k_F} d^3 \vec{k} \sqrt{k^2+m_*^2} \\
p_{st} & = & \frac{1}{3} \frac{2}{(2 \pi)^3} \int_0^{k_F} d^3 \vec{k} \frac{k^2}{\sqrt{k^2+m_*^2}}
\end{eqnarray}
where the Fermi wave number $k_F$ is related to the fermion number density $n$ as follows \cite{basic}
\be
n = \frac{k_F^3}{3 \pi^2}
\en
The integrals above can be computed exactly, and therefore one can obtain analytical expressions for the pressure and energy density of an ideal Fermi gas as follows \cite{thesis}
\begin{widetext}
	\be
	\begin{split}
	\epsilon_{st}  =  \frac{m_*^4}{8 \pi^2} \left((x_F+2x_F^3) \sqrt{1+x_F^2}-sinh^{-1}(x_F) \right) \qquad
	p_{st}  =  \frac{m_*^4}{24 \pi^2} \left((-3x_F+2x_F^3) \sqrt{1+x_F^2}+3 sinh^{-1}(x_F) \right)
	\end{split}
	\en
\end{widetext} 
where we have defined $x_F=k_F/m_*$.

\begin{figure}[ht!]
	\centering
	\includegraphics[scale=0.85]{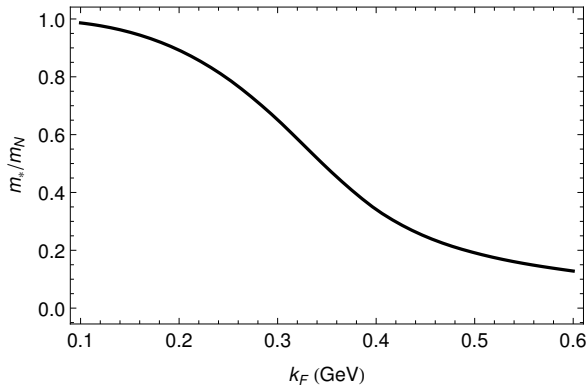}
	\caption{Neutron effective mass versus wave number (in GeV) without dark matter in the $\sigma-\omega$ model.}
	\label{fig:1} 	
\end{figure}

Furthermore we define the scalar baryon density
as follows
\be
n_s = \frac{\partial \epsilon_{st}(m_*)}{\partial m_*}=\frac{2}{(2 \pi)^3} \int_0^{k_F} d^3 \vec{k} \frac{m_*}{\sqrt{k^2+m_*^2}}
\en
to be useful later on, and it is given by
\be
n_s=\frac{m_*^3}{2 \pi^2} \left[ x_F \sqrt{1+x_F^2}-ln\left(x_F+\sqrt{1+x_F^2} \right) \right]
\en
Finally the constant values of the mesons are given by \cite{thesis}
\begin{eqnarray}
V_0 & = & \frac{g_v n}{m_\omega^2} \\
\phi_0 & = & \frac{g_s n_s(m_*)}{m_s^2}
\end{eqnarray}
where the scalar density $n_s$ is evaluated at the neutron effective mass $m_*$. The expression for $\phi_0$ can be obtained from the thermodynamic argument that a closed, isolated system will minimize its energy with respect to the field or the effective mass.
The equation of state is known after the last equation is solved numerically for $\phi_0$. The neutron effective mass as a function of the Fermi wave number can be seen in Fig. 1.

\section{The DM effect on the equation of state}

Now we shall discuss the effect of fermionic DM trapped inside the neutron star on the
nucleon equation of state. Compared to the previous case, we now introduce the DM particle $\chi$ with mass $M$, the Higgs boson $h$ with mass $M_h=125 GeV$, a DM-Higgs Yukawa coupling $y$, and a nucleon-Higgs Yukawa coupling $f m_N/v$, with $v=246 GeV$ is the Higgs vacuum expectation value and $f$ parameterizes the Higgs-nucleon coupling. A complete expression for the factor $f$ can be found e.g. in \cite{singlet3}. Following the lattice computations \cite{lattice1,lattice2,lattice3} we shall consider the central value $f=0.3$ in agreement with \cite{singlet3}. Therefore the system is described by the Lagrangian density
\begin{widetext}
	\be
	\begin{split}
	\mathcal{L}=\mathcal{L}_{had}+\bar{\chi} (i \gamma_\mu \partial^\mu-M+y h) \chi+\frac{1}{2} \partial_\mu h \partial^\mu h-V(h)+\frac{fm_N}{v}\bar{\psi}h\psi
	\end{split}
	\en
\end{widetext} 
where $V(h)$ is the Higgs potential containing the Higgs mass term as well its self-interactions
\be
V(h)=\frac{1}{2} M_h^2 h^2-\lambda v h^3-\frac{1}{4} \lambda h^4
\en
The model is characterized by two free parameters, $M,y$. Having in mind the supersymmetric lightest neutralino \cite{susy1} as the archetypic example of fermionic DM, we can consider a DM mass $M=200 GeV$, while for the DM-Higgs coupling we use the expression (in the large Higgs mixing angle limit) \cite{susy2}
\be
y = \frac{1}{2} (g_2 N_{12}-g_1 N_{11}) N_{14}
\en
where $g_1,g_2$ are the gauge coupling constants of the electroweak sector of the Standard Model, while $N_{1i}$ are the composition coefficients of the lightest neutralino \cite{susy1,susy2}
\be
\tilde{\chi}_1 = N_{11} \tilde{B}+N_{12} \tilde{W}+N_{13} \tilde{H}_u+N_{14} \tilde{H}_d
\en
If $N_{11} \gg N_{12},N_{14}$ the lightest neutralino is Bino-like, if $N_{12}$ dominates the neutralino is wino-like, while if $N_{13},N_{14}$ dominate
the neutralino is higgsino-like. It is easy to construct supersymmetric models with non-universal gaugino mass parameters $M_1,M_2,M_3$ without spoiling naturalness.
The only requirement from LHC searches is that $M_3 \sim m_{\tilde{g}} \geq 1.3 TeV$,
with $m_{\tilde{g}}$ being the gluino mass, while $M_1,M_2$ remain relatively unconstrained \cite{baer}.
Depending on the values of the parameters the DM-Higgs coupling takes values in the range $0.001-0.1$, and in the following we shall adopt the value $y=0.07$.

Finally, we assume that inside the neutron star the DM average number density is $\sim 1000$ times smaller than the average neutron number density, which implies a DM mass fraction $M_{DM}/M_* \simeq 1/6 \simeq 0.17$ \cite{chinos}, with $M_*$ being the mass of the star.

Applying the mean-field approximation to this model, the system looks like an ideal
Fermi gas consisting of two non-interacting fermions with effective masses
\begin{eqnarray}
m_*^n &=& m_N-g_s \phi_0 - f h_0 \\
M_*^\chi &=& M-y h_0
\end{eqnarray}
for neutrons and DM particles respectively, with $h_0$ being the constant value of the Higgs boson
\be
h_0=\frac{f n_s(m_*^n)+y n_s(M_*^\chi)}{M_h^2}
\en
neglecting the Higgs self-interactions in its equation of motion, an approximation that
is verified in the end of the numerical computation given the smallness of the values of the field. The constant value of the vector meson $V_0$ remains the same as before, while $\phi_0$ is given by
\be
\phi_0 = \frac{g_s n_s(m_*^n)}{m_s^2}
\en
Therefore the total pressure and energy density of the system are given by
\begin{widetext}
	\be
	\begin{split}
	p  =  p_{st}(m_*^n)+p_{st}(M_*^\chi) - \frac{m_s^2 \phi_0^2}{2} + \frac{m_\omega^2 V_0^2}{2} - \frac{M_h^2 h_0^2}{2}\qquad
	\epsilon  = \epsilon_{st}(m_*^n) + \epsilon_{st}(M_*^\chi)+\frac{m_s^2 \phi_0^2}{2} + \frac{m_\omega^2 V_0^2}{2}+\frac{M_h^2 h_0^2}{2}
	\end{split}
	\en
\end{widetext}  
The equation of state is known after the system of coupled algebraic equations for $\phi_0,h_0$ are solved numerically. 

\begin{figure}[ht!]
	\centering
	\includegraphics[scale=0.75]{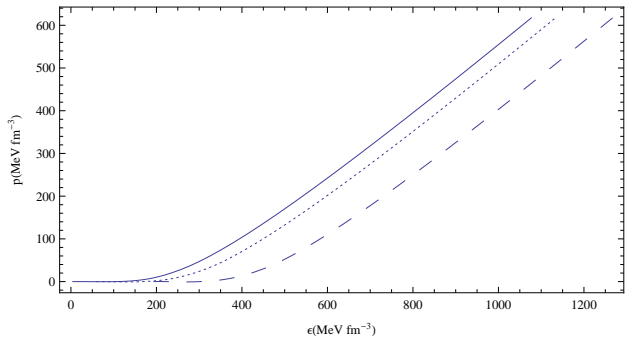}
	\caption{Equation of state of the system neutron-DM for two different values of the DM wave number, $k_F^\chi=0.04 GeV$ (dotted) and $k_F^\chi=0.06 GeV$ (dashed). For comparison the equation of state without DM (solid curve) is also shown.}
	\label{fig:2} 	
\end{figure}

Our main results are shown in figures 2 and 3. In Fig. 2 we show the equation of state, namely pressure versus energy density, for two different values
of the DM wave number $k_F^\chi=0.04 GeV$ (dotted curve) and $k_F^\chi=0.06 GeV$ (dashed curve). For comparison the equation of state without DM (solid curve) is also shown. We see that DM softens the EoS, and therefore one can expect a lower highest neutron star mass in the mass-radius diagram compared to the purely
hadronic case without DM. Indeed, Fig. 3 shows the radius $R$ of the star (in meters) as a function of the star mass $M_*$ (in solar masses) for $k_F^\chi=0.035 GeV$
(lower curve in black). For comparison, we show in the same plot the mass-to-radius relation for pure neutron matter (upper curve in blue). As already anticipated, the admixed DM-neutron star can support a lower highest star mass. In the case of pure neutron star the highest star mass is found to be $M_{max}^{HAD}=2.83 M_{\odot}$, while in the case of admixed DM-neutron star it is found to be $M_{max}^{DM}=2.46 M_{\odot}$. Furthermore, for a given star mass a compact object made of both hadronic matter and DM is considerably smaller than a compact object made of nuclear matter only. Since only nuclear matter can emit thermal radiation, the detection of a compact star with a thermally radiating surface of such a small size could provide a strong evidence for the existence of DM-admixed neutron stars.

A final remark is in order here. The mass-radius diagram has been obtained for non-rotating neutron stars by integrating the TOV equations \cite{TOV} numerically
\begin{eqnarray}
m'(r) & = & 4 \pi r^2 \epsilon(r) \\
p'(r)  & = &  -(\epsilon(r)+p(r)) \frac{m(r)+4 \pi p(r) r^3}{r^2 (1-2m(r)/r)}
\end{eqnarray}
where $r$ is the radial distance, and the prime denotes differentiation with respect to $r$. To find stable configurations for the interior problem $0 < r < R$ we impose initial conditions $m(r=0)=0, p(r=0)=p_c$, with $p_c$ being the central pressure, and require the following conditions on the surface of the star
\begin{eqnarray}
m(r=R) & = & M_* \\
p(r=R) & = & 0
\end{eqnarray}
For rotating neutron stars the highest allowed star mass will be lower by a factor that depends on the rotation speed, but quite generically one expects a decrease of 10 percent roughly. It would be interesting to see for which range of the parameters of the dark matter sector, such as percentage of DM composition etc, the EoS curve does not intersect with the observed bands and therefore would be ruled out. We hope to be able to address this issue in a future work.

\begin{figure}[ht!]
\centering
\includegraphics[scale=0.80]{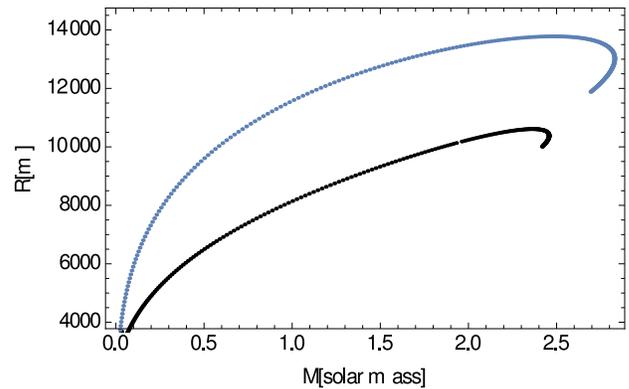}
\caption{Radius (in meters) versus star mass (in solar masses) for pure neutron star (upper curve in blue) and admixed DM-neutron star for $k_F^\chi=0.035 GeV$ (lower curve in black). In the latter case the highest star mass is lower as anticipated.}
\label{fig:3} 	
\end{figure}

\section{Conclusions}

In the present work we have studied the dark matter effect on realistic equation of state of neutron matter in neutron stars. We have assumed that the DM particle is a fermion, and it talks to the Standard Model particles through the SM model Higgs boson. For realistic values of the parameters of the model, namely the DM particle mass and its Yukawa coupling to the SM Higgs boson, we have applied well-established techniques from relativistic mean-field theory models, and we have computed the equation of state of the system numerically. We have compared it with the standard hadronic equation of state without DM, and the mass-to-radius relation for both equations of state (pure neutron matter as well as admixed DM-neutron star) has been shown. Finally, the highest star mass for both cases has been reported.


\begin{acknowledgments}
We would like to thank the anonymous reviewer for valuable comments and suggestions. 
The authors thank the Funda\c c\~ao para a Ci\^encia e Tecnologia (FCT), Portugal, for the financial support to
the Multidisciplinary Center for Astrophysics (CENTRA),  Instituto Superior T\'ecnico,  Universidade de Lisboa,  through the
Grant No. UID/FIS/00099/2013. We wish to thank V. Sagun for discussions on the
mass-to-radius relation of compact objects, and also O. Ivanytskyi for helping us with the radius-mass figure as well as for
enlightening discussions on relativistic mean-field theory models.
\end{acknowledgments}


\end{document}